\documentclass[twocolumn,pre,showpacs,superscriptaddress]{revtex4}
\pdfoutput=1 
\usepackage{graphicx}%
\usepackage{color} 
\usepackage{transparent}
\usepackage{psfrag}
\usepackage{dcolumn}
\usepackage{amsmath}
\usepackage{amssymb}
\usepackage{latexsym}
\usepackage{hyperref} 
\usepackage{booktabs}
\usepackage{latexsym}
\begin{document}

\def\bea{\begin{eqnarray}}
\def\eea{\end{eqnarray}}
\def\beq{\begin{equation}}
\def\eeq{\end{equation}}
\def\f{\frac}
\def\k{\kappa}
\def\e{\epsilon}
\def\be{\beta}
\def\D{\Delta}
\def\th{\theta}
\def\t{\tau}
\def\a{\alpha}
\def\J{{\cal J}}

\def\cDa{{\cal D}[X]}
\def\cDd{{\cal D}[X^\dagger]}
\def\cL{{\cal L}}
\def\cLo{{\cal L}_0}
\def\cLa{{\cal L}_1}

\def\Re{{\rm Re}}
\def\sj{\sum_{j=1}^2}
\def\rk{\rho^{ (k) }}
\def\rek{\rho^{ (1) }}
\def\cek{C^{ (1) }}
\def\rz{\rho^{ (0) }}
\def\rt{\rho^{ (2) }}
\def\rtb{\bar \rho^{ (2) }}
\def\trk{\tilde\rho^{ (k) }}
\def\trek{\tilde\rho^{ (1) }}
\def\trz{\tilde\rho^{ (0) }}
\def\trt{\tilde\rho^{ (2) }}
\def\r{\rho}
\def\tD{\tilde {D}}
\def\C{{\cal {C}}}

\def\s{\sigma}
\def\kb{k_B}
\def\F{{\cal F}}
\def\la{\langle}
\def\ra{\rangle}
\def\nn{\nonumber}
\def\up{\uparrow}
\def\dn{\downarrow}
\def\S{\Sigma}
\def\dg{\dagger}
\def\d{\delta}
\def\p{\partial}
\def\l{\lambda}
\def\le{\left}
\def\ri{\right}
\def\L{\Lambda}
\def\G{\Gamma}
\def\o{\Omega}
\def\w{\omega}
\def\g{\gamma}
\def\mt{\tilde {m} }
\def\mbt{\tilde {\bf m} }

\def\rv{ {\bf r}}
\def\jv{ {\bf j}}
\def\jr{ {\bf j}_r}
\def\jd{ {\bf j}_d}
\def\noi{\noindent}
\def\a{\alpha}
\def\d{\delta}
\def\p{\partial} 

\def\H{ {\bf H}}
\def\He{{\bf H_e}}
\def\h{{\bf h}}
\def\m{{\bf m}}
\def\hth{h_{\theta}}

\def\Dn{D_{\alpha} }
\def\Rp{R_{\parallel}}
\def\Ng1{N_d^{(1)}}
\def\la{\langle}
\def\ra{\rangle}
\def\e{\epsilon}
\def\n{\eta}
\def\g{\gamma}
\def\break#1{\pagebreak \vspace*{#1}}
\def\hf{\frac{1}{2}}

\title{Confinement and crowding control the morphology and dynamics of a model bacterial chromosome} 
\author{Pinaki Swain}
\email{ch13p1002@iith.ac.in}
\affiliation{Indian Institute of Technology Hyderabad, Kandi, Sangareddy 502285, Telangana, India}
\author{Bela M. Mulder}
\email{mulder@amolf.nl}
\affiliation{Institute AMOLF, Science Park 104, 1098XG Amsterdam, Netherlands}
\author{Debasish Chaudhuri}
\email{debc@iopb.res.in}
\affiliation{Institute of Physics, Sachivalaya Marg, Bhubaneswar 751005, India
}
\affiliation{Homi Bhabha National Institute, Anushaktinagar, Mumbai 400094, India}

\date{\today}

\begin{abstract}
Motivated by recent experiments probing shape, size and dynamics of bacterial chromosomes in growing cells, we consider a polymer model consisting of a circular backbone to which side-loops are attached, confined to a cylindrical cell. Such a model chromosome spontaneously adopts a helical shape, which is further compacted by molecular crowders to occupy a nucleoid-like subvolume of the cell. With increasing cell length, the longitudinal size of the chromosome increases in a non-linear fashion to finally saturate, its morphology gradually opening up while displaying a changing number of helical turns. For shorter cells, the chromosome extension varies non-monotonically with cell size, which we show is associated with a radial- to longitudinal spatial re-ordering of the crowders. Confinement and crowders constrain chain dynamics leading to anomalous diffusion. While the scaling exponent for the mean squared displacement of center of mass grows and saturates with cell length, that of individual loci displays broad distribution with a sharp maximum.
\end{abstract}


\maketitle


\section{Introduction}
The bacterial chromosome of \emph{E. coli} consists of a $1.6\,$mm long negatively supercoiled circular 4.6 Mbp DNA-strand~\cite{Wang2013}. Unlike eukaryotes, bacteria do not have a separate membrane-bound nucleus, so their chromosome is suspended in the cytoplasm and only confined by the cell envelope.  A wild type {\em E. coli} cell has a roughly cylindrical shape, with diameter $0.8\,\mu$m and length varying between 2 - 4$\,\mu$m. 

Within the cell volume the nucleoid is observed to occupy a subvolume of dimensions approximately $0.5\,\mu$m by $1.6\,\mu$m. This implies that the chromosome is compacted roughly three orders of magnitude in volume, compared to its freely expanded state~\cite{Fisher2013}. A number of effects have been discussed as contributing to this compaction: the physical confinement to the cellular volume, DNA supercoiling, and depletion effects caused by the molecularly crowded cytosol. Moreover, contour-wise distant  parts of the DNA are brought into spatial proximity leading to loop structures~\cite{Worcel1972, Holmes2000, Zimmerman2006}. A number of proteins bind and stabilize these loops into separate topological domains \cite{Brocken2018, Goloborodko2016,Song2015}. Direct morphological evidence for loop structures were first observed in electron microscopy (EM) experiments~\cite{Delius1974, Kavenoff1976,  Trun1998}. Later EM studies on lysed {\em E. coli} bacteria showed a bell-shaped distribution of loop sizes with a maximum close to 10-12 kbp~\cite{Postow2004}. These results are complemented by contact maps, showing spatial contacts of different genes along the DNA contour~\cite{Le2013, Marbouty2014}. 

In remarkable recent experiments, live cell imaging revealed that nucleoids in rod shaped bacteria like {\em E. coli}~\cite{Fisher2013} and {\em B. subtilis}~\cite{Berlatzky2008} exhibit a ubiquitous helical shape with pitch-length a fraction of the cell length. Further, inhibiting cell wall formation locally rounded up the cell, modifying chromosome helicity~\cite{Fisher2013}, showing impact of cellular confinement on  chromosomal morphology. Live cell imaging also showed anomalous diffusion for chromosome loci in {\em E. coli} and {\em C. cresentus}~\cite{Weber2010, Kuwada2013, Javer2014, Lampo2015}.   

In this paper we explore to what extent such emergent properties may have purely physical origin using computer simulations on a model bacterial chromosome. The bacterial cytosol is a high density poly-dispersed material, which is known to be fluidised by metabolic activity~\cite{Parry2014}. The strong cellular confinement introduces hydrodynamic screening, reducing the impact of hydrodynamic coupling in the dynamics. We therefore model the cellular environment as a free flowing fluid represented by a Langevin heat bath. Although, local structure of a chromosome is expected to involve distribution of loop sizes and topological entanglement, we consider  arguably the simplest  model of a ring polymer to which closed side loops of equal size are attached with a uniform spacing~\cite{Reiss2011, Chaudhuri2012, Chaudhuri2018}. The effect of these side-loops can be captured through an excess Gaussian core repulsion between the backbone beads~\cite{Chaudhuri2012}. Such a polymer spontaneously organizes into a helicoid morphology when confined to a cylindrical cell-like volume. 

As was suggested previously, crowding effects due to cytosolic protein(complexes) could further compact the chromosome to a sub-volume of the cell~\cite{Odijk1998, Woldringh1995}. Simulations involving a model plectonemically folded chromosome indeed showed expulsion of crowders representing transcribing ribosomes and mRNA to the cell poles \cite{Mondal2011}, recapturing nucleoid morphologies observed in live {\em E. coli} cells~\cite{Chai2014, Nevo-Dinur2011}. We include them in the form of non-additive crowders within our model.

We first study how the size of the chromosome responds to changes in cell length. In longer cells, chromosome extension increases monotonically with cell length in a non-linear fashion, finally saturating at an extension that remains significantly smaller than the full extension possible in absence of crowders. The change in slope of the chromosome extension with the length of a growing cell is governed by the crowder density. This behaviour is well captured by a simple model based on scaling arguments. However, for the strongly confined chromosomes at smaller cell sizes, the extension exhibits a remarkable non-monotonic variation associated with radial to longitudinal re-organization of crowders. We present a mean field approach to explain this behavior. The compression due to crowders not only determines the chromosome size, but, as we show, it also maintains the helicoid shape of the chromosome even in very long cells, with lengths that exceed the full extension of the chromosome by multiple times. With increasing cell length, the number of helical turns of the chromosome reduces to finally saturate, where the saturation value is determined by the crowder density. Finally, turning to dynamical effects, we observe that monomers show anomalous diffusion with a narrow distribution of exponents having values similar to that in live cells~\cite{Weber2010}. The chromosome center of mass also shows anomalous diffusion but with an exponent that increases with the cell length, but saturates in a manner observed in recent experiments on live {\em E. coli} cells~\cite{Wu2018}. 

\section{Model}
Our model coarse-grained chromosome consists of a circular backbone chain of $n_m$ beads of diameter $\s$, which sets the bond length, and hence of total length $\ell_m=n_m \s$.  To each backbone bead we connect a closed side-loop of of $n_s$ beads, and hence length $\ell_s=n_s \s$. The total length of the chromosome is thus $\ell=\ell_m +  n_m\, \ell_s$. 
The bonded-beads interact via a finitely extensible non-linear elastic (FENE) \cite{Kremer1990} potential  $\be V_{b} = - \hf K R^2 \ln \le[ 1 - {\le( \f{d_i-\s}{R} \ri)}^2 \ri]$, where $\beta = 1/\kb T$ is the canonical inverse temperature, ${\bf d_i} = {\bf r_{i+1}} - {\bf r_i}$ is the bond vector, with bond length $d_i=|{\bf d_i}|$. The parameters  $K=30$, and $R=1.5 \s$ determine bond fluctuations. The self-avoidance between non-bonded beads is modelled via a short-ranged repulsive Weeks-Chandler-Andersen (WCA) potential~\cite{Weeks1971}, 
$
\beta V(r_{ij}) = 4  [ (\s/r_{ij})^{12} - (\s/r_{ij})^6] +1/4
$
when inter-monomer separation $r_{ij}< 2^{1/6} \s$, $\be V(r_{ij})=0$ otherwise.  
The WCA and FENE potential together set the bond length to  $b=1.09\,\s$.  

Previous all-atom simulations of the above model have been used to extract the effective repulsion between side-loops. It was shown to be well represented by a soft Gaussian core potential $\be V_{gc} (r)= a \exp[-r^2/2 \S^2]$, with $\S^2 \approx  2 R_g^2$, where the radius of gyration of each loop $R_g = c n_s^{3/5} \s$ with $c=0.323$~\cite{Chaudhuri2012, Bolhuis2001, Narros2013}. The values of $a$ and $\S$ depends on the side loop size, and discussed later in this section.

Confinement is imposed by implementing a repulsive interaction between all beads and the surfaces representing the cell volume. The latter we model as a straight cylinder of length $L$ and diameter $D$. The specific form of the wall-repulsion was chosen as
$
\be V_{\rm wall} = 2\pi\e [(2/5)(\s/r_{iw})^{10}-(\s/r_{iw})^4 + 3/5]
$
when the distance of the $i$-th monomer from a wall $r_{iw} < \s$ and $\be V_{\rm wall} =0$ otherwise. In addition, a Gaussian core repulsion with the walls $\be V_{gc}(r_{iw})$ with $\S=R_g$ and strength $a/2$ is considered to model repulsion from the wall due to side-loops.  

To model the depletion effects due to the molecularly crowded cytosolic environment, we introduce so-called non-additive crowder particles, which do not interact among themselves but repel chromosomal beads via a combination of the WCA potential $\beta V(r_{ij})$ and Gaussian core $\be V_{gc}(r_{ij})$ where $i$ denotes a crowder and $j$ a polymer bead~\cite{Dijkstra1998}. Their repulsion from the walls is also taken as a combination of  $\be V_{\rm wall} $ and $\be V_{gc}(r_{iw})$, i.e., the same as that between beads and walls. We assume that a growing cell will maintain a constant density of depletants $\r$, so that the number of crowders as a function of the cell length is given by $N_d = \r \f{\pi}{4}D^2 L$. As translation of proteins occurs primarily near the chromosome in the cytosol~\cite{Kuhlman2012}, we introduce new crowders when needed using spatially uniform insertion attempts around the model chromosome with acceptance based on the Widom criterion, i.e., proportional to the change in Boltzmann weight caused by the proposed insertion~\cite{Binder1997,Widom1963}.

\begin{figure}[t] 
\begin{center}
\includegraphics[width=8cm]{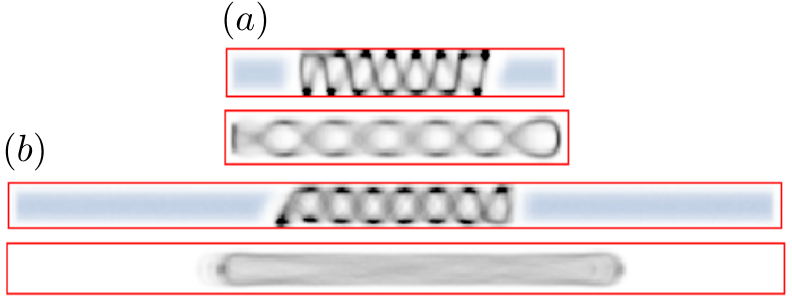}
\caption{(Color online) Local density plots of the model chromosome backbone in the presence or absence of crowders in a cell of  $L=9\,D$\,($a$) and $L=21\,D$\,($b$). The density of crowders (top of each pair) is $\r D^{3}= 0,\, 212$.
The local crowder density is shown by the bluish gray shade. The white area within the polymer, and that between the polymer and crowders denotes the volume excluded by side-loops. Note that in the $L=9\, D$ cell with $\r D^{3}= 0$ the chromosome expands to touch both the cell caps.}
\label{fig:config1}
\end{center}
\end{figure}

Throughout, we adopt a fixed cell diameter of $D = 26.67 \s$, while the cell length varies between $3\text{--}30\, D$. To model a Mbp long bacterial chromosome consisting of 10\,kbp long side loops, we use a backbone of $n_m=636$ beads, incorporating side-loops of $n_s=62$ beads.  Each side loop has a radius of gyration $R_g = c n_s^{3/5} \s = 0.14 \, D$,  which is much smaller than the diameter of the cell $D$. While it is well known that effective repulsion between polymers in free space is independent of the polymer size~\cite{Cloizeaux1975, Daoud1975, Grosberg1982}, in cylindrical confinement the blob picture~\cite{Gennes1979} suggests that the repulsion strength is expected to be proportional to the polymer length, provided the polymer under consideration is long and breaks into multiple blobs in the given confinement~\cite{Jun2006}. As the side-loops we consider have $R_g \ll D$, the repulsion between them behaves like that in bulk. Some of the side loops may get into a chain-like plectoneme conformation, locally maintained by supercoiling. 
The effective repulsion strength between long open chains is known to be $a = 2$~\cite{Bolhuis2001}, while that between loops is a weak function of loop size and detailed topology, and remains within the range $2 < a<6$ in bulk~\cite{Narros2013}.  To incorporate possibilities of both loop and plectoneme morphology of side-loops, in all our simulations we fix the value of inter-loop repulsion $a$ to an intermediate value, $a=3$, and  $\S = \sqrt 2 R_g = 0.2\, D$. 

We simulate this system employing a velocity-Verlet molecular dynamics (MD) scheme in the presence of a Langevin thermostat fixing the temperature at $\kb T =1$,  as implemented by the ESPResSo package~\cite{Limbach2006}. Diffusion of a single particle over the cell diameter $\s$ takes $\t = \s^2 (\a/\kb T)$, where $\a = 3\pi \eta \s$ is the viscous dissipation due to the cytosol of viscosity $\eta$. The time scale is set by $\t$. Integration is performed over time step $\d t = 0.01\t$.
In discussing simulation results we present all length scales in units of $D$ and time scales in units of $\t$.
All results are presented analyzing $10^3$ configurations after equilibration.  These are separated by $10^4$ time steps to avoid possible bias due to temporal correlations. 

\begin{figure}[t] 
\begin{center}
\includegraphics[width=8cm]{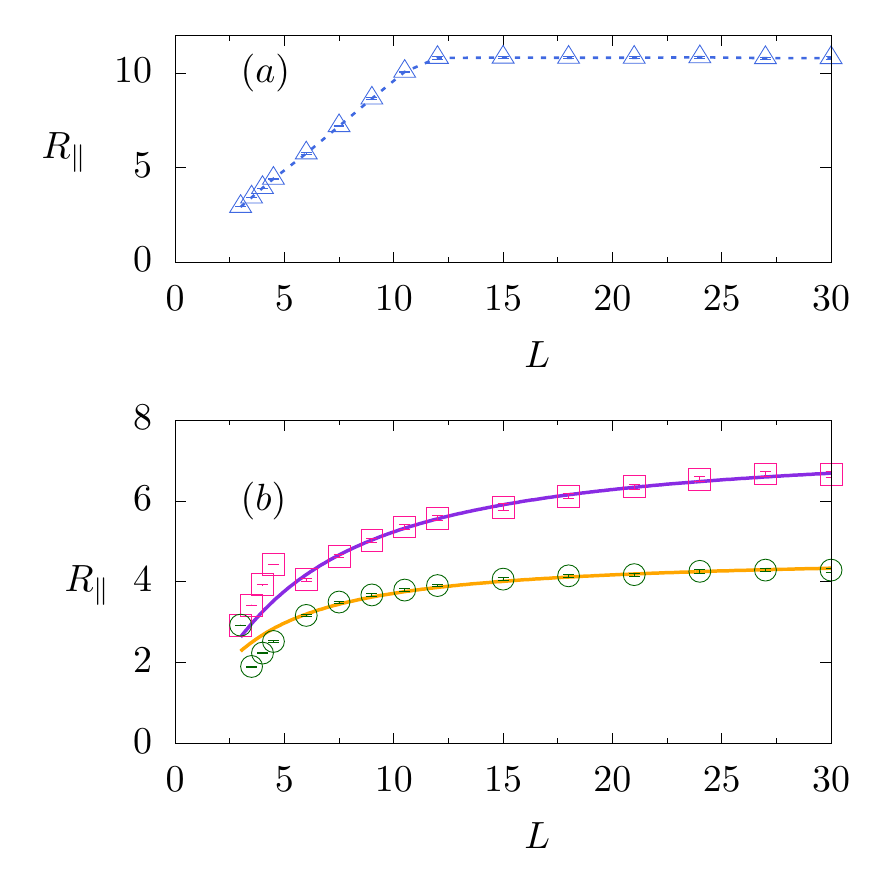}
\caption{(Color online) Longitudinal extension $\Rp$ of chromosome as a function of $L$, ($a$)\,in absence of crowders, ($b$)\,in presence of crowders. 
Symbols $\bigtriangleup$, $\Box$, $\bigcirc$ denote data at crowder densities $\r D^{3}=0,\,212,\, 1060\, $ 
respectively.  In ($b$), the solid lines show fit of data to Eq.(\ref{eq:rp_m}) with $a=4.9\times 10^{-5}\,D^{-2}$, and 
$b=1.22 \times 10^4 \, D$ for $\r D^{3}=212$,  $b=2.33 \times 10^4\, D$ for $\r D^{3}=1060$.
All length scales are expressed in units of $D$. 
}
\label{fig_rpll}
\end{center}
\end{figure}

\section {Chromosome size and morphology}

For smaller cell lengths we observe, both in the presence or absence of crowders, the spontaneous emergence of helicity in the backbone of the polymer. The crowders are expelled from the polymer and intercalate symmetrically between the bulk of the polymer and the end walls of the cylinder. In the absence of crowders the backbone can expand to such a degree that it touches the end walls. These effects are illustrated in Fig.~\ref{fig:config1}($a$), which depicts the density of the polymer backbone and crowders inside a cylindrical cell of length $L=9\, D$. 
In longer cells the differences are more marked. Without crowders the chromosome can basically fully expand and ``unwind'', while in the presence of crowders the chromosome remains compacted due to the depletion forces and retains its helicity. This is illustrated in Fig.~\ref{fig:config1}($b$), which depicts the chromosome in a cell of length $L=21\,D$. 

\subsection{Size}
Chromosome size shows a non-linear dependence on the degree of confinement, the detailed nature of which gets modified with increasing density of crowders.
\subsubsection{In absence of crowders}
The chain size $\Rp$, defined as the longest extension of the polymer backbone along the cylinder axis, increases linearly up to the cell length $L = \Rp^{\text{max}}$, beyond which it ceases to grow further~(Fig.~\ref{fig_rpll}($a$)). 
As we show later in Sec.\,\ref{mft}, the maximal extension of chromosome follows the de Gennes' scaling of a self avoiding chain in an infinite cylinder $\Rp^{\text{max}} \sim N D^{-2/3}$. For further quantification of chromosome size in terms of radii of gyration, see Appendix-\ref{no_crowder}  

\begin{figure}[t] 
\begin{center}
\includegraphics[width=8cm]{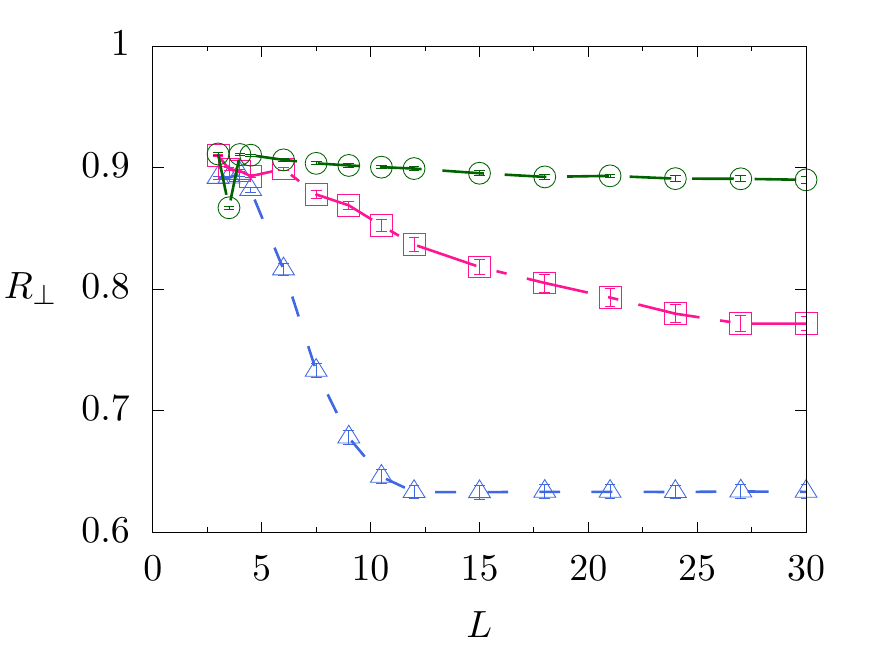}
\caption{(Color online) Transverse size of chromosome $R_\perp$ decreases as a function of cell length $L$. 
Symbols $\bigtriangleup$, $\Box$, $\bigcirc$ denote data at crowder densities $\r D^{3}=0,\,212,\, 1060\, $ 
respectively.  
All length scales are expressed in units of $D$. 
}
\label{fig_rperp}
\end{center}
\end{figure}

\subsubsection{In presence of crowders}

In presence of crowders, the variation of chromosome size is qualitatively different from the case without, discussed above. 
For small cell lengths, we observe a non-monotonic variation of $\Rp$ with $L$~(Fig.\ref{fig_rpll}($b$)). This persists at all crowder densities $\r$, but the point of non-monotonic jump gets suppressed towards smaller $L$ values with increase in $\r$. We return to this point with a detailed explanation in Sec.\ref{sec:reconf}.

In longer cells, the chromosome length increases gradually with cell length, deviating from linearity almost immediately and smoothly reaching a saturation value~(Fig.\ref{fig_rpll}($b$)). Moreover, the maximum extension of chromosome remains significantly smaller than that in absence of crowders, its value strongly controlled by the crowder density.  These features can be understood semi-quantitatively using the mean field argument we present below. Fig.\ref{fig_rpll}($b$) shows a nice fit of simulation data with the mean field prediction. Moreover, the behavior observed in Fig.\ref{fig_rpll}($b$)  captures qualitatively the variation of extension of {\em E. coli} chromosome in growing cells~\cite{Wu2018}. Finally, we note that associated with the increase in $\Rp$ as the cell length $L$ grows, the radial size $R_\perp$ - measured in terms of corresponding component of radius of gyration tensor - decreases to finally saturate~(Fig.\ref{fig_rperp}), a behavior typical of non-auxetic materials.

\subsubsection{Mean field estimate}
\label{mft}
The free energy of a self avoiding chain in a cylinder may be written down in terms of the de Gennes' blob picture. Assume that the polymer confined in a cylinder  of diameter $D$, consists of $N/g$ blobs where each blob contains $g$ monomers out of $N$ available in the chain. Expressing the size of the polymer along the long axis of the cylinder $\Rp$, the  free energy of the chain is expressed as 
\bea
\be F_c = A_1 \f{\Rp^2}{(N/g) D^2} + B_1 \f{D(N/g)^2}{\Rp}. \nn
\eea  
The three-dimensional Flory scaling is maintained within a blob, $g \sim D^{5/3}$.  Using this in the above expression,
\bea
\be F_c = A \f{\Rp^2}{N D^{1/3}} + B \f{N^2}{D^{7/3} \Rp}. 
\eea
We use this free energy to describe the effective chromosomal backbone.    
Assuming that the crowders are segregated longitudinally from the chromosome, the volume occupied by the crowders is $V_d =C_d D^2(L-\Rp)$, while the rest of the volume is occupied by the chromosome. The geometrical prefactor $C_d=1$ for rectangular parallelepiped, and $C_d=\pi/4$ for cylinders.  The free energy of the non-additive crowders is 
\bea
\be F_d = N_d \left[ \ln\f{ N_d \s^3}{V_d} - 1 \right].
\label{eq:fd}
\eea 

Minimizing the total free energy $F_t = F_c+F_d$ one obtains a quartic equation in $\Rp$,
\bea
\f{\p \be F_t}{\p \Rp} = \f{2A \Rp}{N D^{1/3}} - \f{B N^2}{D^{7/3} \Rp^2} + \f{N_d}{C_d(L-\Rp)} = 0. 
\label{eq:rp_m}
\eea

For cells with constant density of crowders, one replaces $N_d=C_d \r D^2 L$ in the above equation. In the limit of large $N,\, N_d \gg 1$, neglecting $a = 2A/N D^{1/3}$ one gets a quadratic equation $N_d \Rp^2 + b \Rp - b L = 0$, with $b = C_d B N^2/ D^{7/3}$ that has a dimension of length. 
This equation has a closed form solution 
\bea
\Rp \approx \left[\sqrt{b^2 + 4 b L N_d} - b \right]/ 2 N_d. 
\label{eq:rp}
\eea
As is evident from the fits in Fig.\ref{fig_rpll}($b$), the limit $a\to 0$ holds well for the two data sets, and Eq.(\ref{eq:rp}) describes these results. 

In the limit of both long $L$ or large $N_d$, this relation simplifies further to $\Rp \approx 2 \sqrt{bL/N_d}$. For constant density of crowders, $N_d = \r \pi D^2 L/4$, one obtains the maximum possible extension in long cells reducing with crowder density as $\Rp \sim \r^{-1/2}$, indicating how the size of a nucleoid-like sub-volume is maintained by the crowders. 
On the other hand, if one considers a situation where the cell length remains unchanged, with increasing $N_d$ one crosses from a regime $N_d \ll b/4L$ in which $\Rp \sim N_d^{-1}$ to $N_d \gg b/4L$ where $\Rp \sim N_d^{-1/2}$. 
This situation is considered in detail in Sec.\ref{sec:Nd}.
A third possibility is to keep $N_d$ fixed in a growing cell such that the overall crowder density keeps decaying. In this case Eq.\ref{eq:rp} suggests a growth in nucleoid size as  $\Rp \sim L^{1/2}$ that will finally saturate to the value of maximal extension of chain in absence of crowders (see Appendix-\ref{fixed_Nd}). 

In Eq.(\ref{eq:rp_m}), the term $N_d/(L-\Rp)$ denotes a linear density of crowders. In the limit of vanishingly small density, the first two terms in Eq.(\ref{eq:rp_m}) leads to the relation
$$(C_d a \Rp^3 -b)(L-\Rp)=0,$$
which, gives $\Rp =(b/C_d a)^{1/3} \sim N D^{-2/3}$ the de-Gennes' scaling form for long cylinders, 
$L >  \Rp$,
and $\Rp=L$ otherwise. 

\subsection{Spatial reconfiguration}
\label{sec:reconf}
In this section we discuss how the non-monotonic variation of $\Rp$ with $L$ can be understood in terms of a spatial reordering between the chromosome and crowders, in a growing cell with constant crowder density. Further, we show how the same effect impacts the chromosome extension in a given confinement with increasing number of crowders.

\subsubsection{With increasing length}
\label{non_mono}
\begin{figure}[t] 
\begin{center}
\includegraphics[width=8cm]{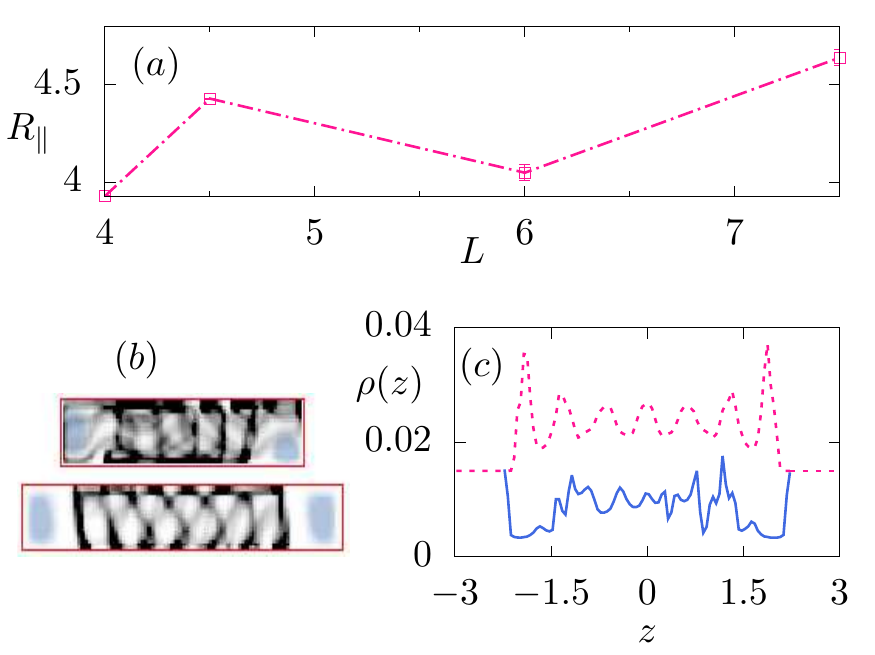}
\caption{(Color online)   ($a$)\,Magnified representation of the non-monotonic variation of $\Rp$ with $L$ at $\r D^3=212$. 
($b$)\,Local density profile of chromosome and crowders at $L= 4.5,\,6$. 
($c$)\,Density profile of chromosome along the longitudinal direction $\r(z)$ for cell lengths $L=4.5,\, 6$ denoted by the blue solid line, and the magenta dashed line, respectively. The graph at $L=6$ is shifted upwards by an additive factor $0.015$ for better visibility. 
All length scales are expressed in units of $D$. 
}
\label{ch_sz_nonmono}
\end{center}
\end{figure}

The variation of the longitudinal size of the chromosome with increasing cell length, at a constant density of crowders, e.g.,  $\r D^3 = 212$, shows a striking non-monotonicity at small cell lengths, e.g., between $L=3\,D$ to 6$\,D$~(Fig.~\ref{fig_rpll}($b$) and Fig.~\ref{ch_sz_nonmono}). 
Such discontinuous and non-monotonic size changes are associated with a reordering of the spatial distribution of crowders relative to the chromosome. This is directly observable from local density plots of chromosome and crowders in Fig.s~\ref{ch_sz_nonmono}($b$) and ($c$) using $10^3$ equilibrium configurations. The apparent overlap of crowder and monomer density in the top panel of $(b)$ is due to the two dimensional projection of the three dimensionally segregated density profiles along radial direction. The narrowing of the linear density of backbone chain along the cell length, $\r(z)$,  with increasing cell size from $4.5 D$ to $6D$ clearly shows an associated compression of the chromosome. 

For smallest cell lengths, entropic repulsion by the chain in longitudinal direction is too strong expelling crowders from the planar caps of the cylinder. In this limit chromosome and crowders are essentially radially segregated near the two caps of the cylinder. Thus the longitudinal extension of the chromosome follows $\Rp=\Rp^0$, where $\Rp^0$ denotes the extension  in absence of crowders, as shown in Fig.~\ref{fig_rpll}($a$).  As the cell length grows, from $L=4.5\,D$ to $6\,D$, the additional space along the cell length allows for relaxation of longitudinal stress due to chromosome and, as a result, a polar relocation of crowders takes place~(Fig.s~\ref{ch_sz_nonmono}($b$),($c$)), compressing the chromosome. This leads to the non-monotonic variation of the nucleoid length, changing from the linear form of $\Rp=\Rp^0 \sim L$ to the approximate non-linear dependence $\Rp \approx [\sqrt{b^2+4 b L^2 \r} -b]/2\r L$. The higher compressional stress at higher crowder density $\r D^3=1060$ shifts the point of non-monotonicity to a smaller cell size~(see Fig.\ref{fig_rpll}).

 \begin{figure}[t] 
\begin{center}
\includegraphics[width=8cm]{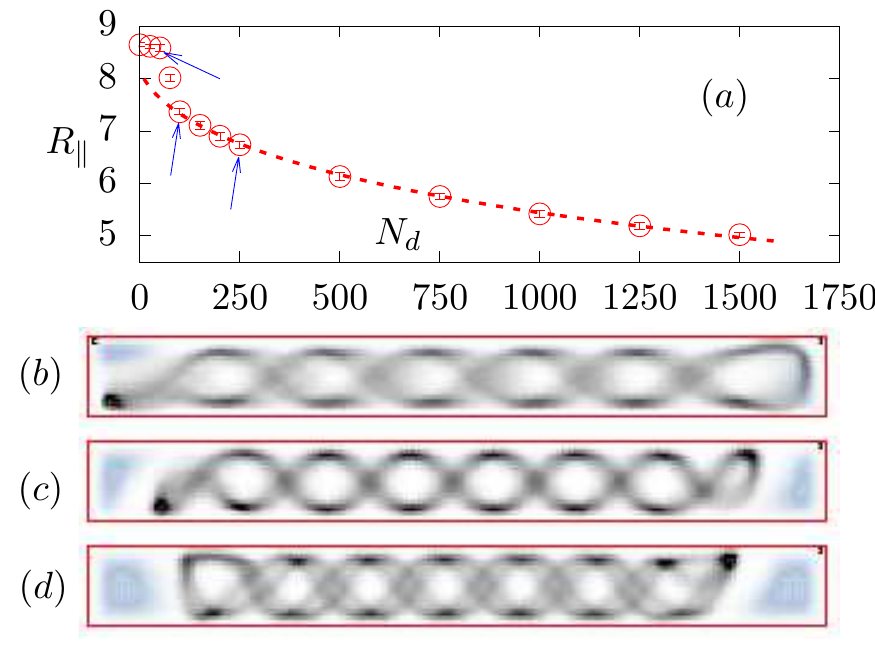}
\caption{(Color online) ($a$)\,Chromosome size $\Rp$ (in units of $D$) as a function of number of depletants $N_d$ in a cell of length $L/D=9$. Data points show simulation results. 
The dotted line  is a fit to Eq.\ref{eq:rp_m}, with fitting constants $a=21.91\,D^{-2}$, $b=1.18 \times 10^4\,D$. 
Arrows in ($a$) indicate the points at which local density plots for polymer and crowder configurations are shown in ($b$) $N_d=50$, ($c$) $N_d=100$, and ($d$) $N_d=250$. 
}
\label{Ree_L3}
\end{center}
\end{figure}

\subsubsection{With increasing density}
\label{sec:Nd}
Similar spatial reconfiguration is observed by varying the crowder density keeping the cell length fixed. In Fig.~\ref{Ree_L3} we show the results for changing the number of crowders $N_d$ in a cell of a given length, $L=9\,D$. One finds a sharp drop in longitudinal chrosmosome size $\Rp$ from a cell- spanning structure to one covering approximately half the cell length as $N_d$ increases from $0$ to $1500$~(Fig.~\ref{Ree_L3}($a$)). At very small densities, $\Rp$ remains almost unchanged with increasing $N_d$.  
Beyond this regime the chromosome extension reduces discontinuously with $N_d$ to finally shrink in a nonlinear fashion that is described by Eq.(\ref{eq:rp_m}). The dotted line in  Fig.~\ref{Ree_L3}($a$) shows a fit to this equation.    

Fig.~\ref{Ree_L3} ($b$)- ($d$) shows local density plots of crowders and the the chromosome using $10^3$ equilibrium configurations. At small crowder density, chromosome extends to almost the full length, to push the crowders away radially to make space at the two caps of the cylinder~(Fig.~\ref{Ree_L3} ($b$)). This is similar to the effect described in Sec.~\ref{non_mono}. The apparent overlap of crowder density and chromosome configuration observed in this figure is due to the two dimensional projection of the original three dimensional organization, in which the monomer and crowder densities segregate predominantly along the radial direction. As the density of crowders increases, they occupy the cylinder caps compressing the chromosome longitudinally~(Fig.~\ref{Ree_L3} ($c$),($d$)), making the helical turns more compact.

\subsubsection{Mean field description}
\label{mft_reorg}
Clearly Eq.(\ref{eq:rp_m}) can not capture the variation of $\Rp$ for all the cell lengths. Particularly, below $N_d=100$ in Fig.\ref{Ree_L3}($a$), the chromosome extension increases sharply to saturate to the the complete extension $\Rp^0$. Such a sharp change is indicative of a phase transition. As Fig.\ref{Ree_L3}($b$),($c$) shows, across the transition, crowders near the two caps undergo a spatial reconfiguration from a longitudinal to radial organization. In deriving Eq.(\ref{eq:rp_m}), we did not incorporate this possibility, and assumed that crowders are always longitudinally segregated from chromosome. The possibility of the radial reordering of crowders may be incorporated in the theory by assuming  that the chain covers only a fraction $\Dn = \alpha D$ of the radial space of the cylinder, whereas $D-\Dn$ is occupied by the crowders. Thus the confinement induced blob-size would change to $g \sim \Dn^{5/3}$. The corresponding free energy of the confined chain is
\bea
\be F_c 
= A \f{\Rp^2}{N \Dn^{1/3}} + B \f{N^2}{\Dn^{7/3} \Rp}.\nn
\eea 
As before, the effect of depletion can be incorporated by accounting for the unavailability of space occupied by the chromosome to crowders. For simplicity, considering the confining geometry as a rectangular parallelepiped, one can express the  free energy due to crowders as the following,
\bea 
\be F_d = N_d \left[ \ln\f{ N_d \s^3}{(D-D_\a)^2 L + D_\a^2(L-\Rp)} - 1 \right].\nn
\eea 
Note that, in the limit of $D_\a \to D$, the above expression reduces to the free energy contribution due to crowders in Eq.(\ref{eq:fd}).
Minimization of the total free energy $(F_C+F_d)$
with respect to $\Rp$ and $\alpha$ simultaneously, would then formally describe the impact of radial to longitudinal reconfiguration on the equilibrium extension of the polymer. 

However, this simple mean field description suffers from the assumption of a uniform radial compression of chromosome to size $\Dn$, while the simulation results clearly showed that the radial compression of chromosome and the radial reconfiguration of crowders and chromosome takes place only near the two caps of the cylinder. Thus, any attempt to compare this theory with simulation results is bound to overestimate the effect of radial segregation. However, in polymer models with much thinner backbone and smaller crowder size, the overall radial segregation between confined crowders and chains can not be preempted. It remains to be seen how this mean field description captures spatial reordering under such conditions.

\begin{figure}[t] 
\begin{center}
\includegraphics[width=8cm]{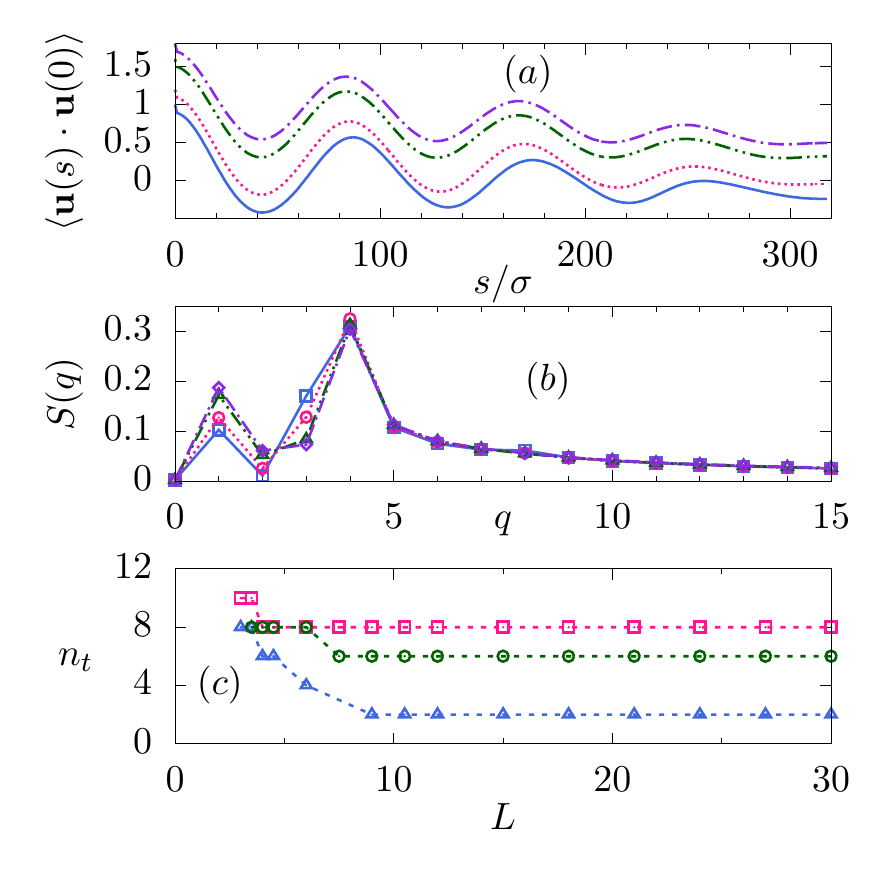}
\caption{(Color online) Helicity of chromosome: ($a$)\,Local tangent ${\bf u}(s)$ correlation function in chromosome backbone from $s=0$ to $\ell_m/2$ at crowder density $\r D^3=212$ for cell lengths $L=9,\, 12,\, 21,\, 27$ denoted by dashed line (blue), dash- dotted line (magenta), big-dashed line (green), small dash-dotted line (indigo) respectively. Each graph corresponding to a larger $L$-value is shifted upwards with respect to that having a smaller $L$ by an additive factor $0.2$ for better visibility. ($b$)~Corresponding structure factors at $L=9\,(\Box),\, 12\,(\bigcirc),\, 21 \, (\bigtriangleup),\, 27\, (\diamond)$ capture the periodicity in correlation, with the peak position $q = q_p\, (>1)$ denoting the number of turns in the emergent helix. ($c$)\,The number of helical turns $n_t = 2 q_p$ is plotted as a function of $L$. Symbols $\bigtriangleup$, $\Box$, $\bigcirc$ denote data at crowder densities $\r D^3=0,\,212,\, 1060$ respectively.   
All length scales are expressed in units of $D$. 
}
\label{ch_helicity}
\end{center}
\end{figure}

\subsection{Morphology}
Even in absence of crowders, a competition between the effective bending rigidity due to side-loops, and their relative packing within the confinement leads to the spontaneous formation of helical morphology, with a number of helical turns organized along the axial direction of the cylinder (see Fig.~\ref{fig:config1}, and also Fig.~\ref{fig:configs_ro1060} in Appendix-\ref{app:morpho}). We can characterize the helicity of the chain through spatial oscillations of the tangent-tangent correlation function $\la {\bf u}(s) \cdot {\bf u}(0) \ra$~(Fig.~\ref{ch_helicity}($a$)), and the corresponding structure factor $S(q)$~(Fig.~\ref{ch_helicity}($b$)). 
The correlation function is shown up to $s=\ell_m/2$, the longest separation along the chain of total length $\ell_m=n_m \s$. 
The position of the dominant peak in structure factor at $q=q_p$ provides the number of helical turns.
The total number of this for the full chain is  $n_t = 2 q_p$.  The magnitude $S(q_p)$ quantifies the degree of helicity. The helical pitch length $\ell_p$ is related to $n_t$ and is given by $\ell_p = \ell_m/n_t$. 
Similar analysis at a higher crowder density $\r D^3=1060$ is presented in Appendix-\ref{app:morpho}.
Fig.~\ref{ch_helicity}($c$) shows  the total number of helical turns  along the full backbone with increasing cell length, corresponding to three different cytosolic densities $\r D^3=0,\,212,\, 1060$.

\subsubsection{In absence of crowders}
A simple geometric analysis of a helical curve of length $\ell$ absorbed on the surface of a cylinder gives an estimate  $n_t \approx (\ell^2 - \Rp^2)^{1/2} / \pi R_\perp$, in terms of the longitudinal extension $\Rp$ and radial size $R_\perp$. Assuming $R_\perp \approx D$ in absence of molecular crowders, and given that $\Rp$ grows monotonically to saturate with increasing $L$, the number of helical turns is expected to decrease as $(\ell^2 - p^2 L^2)^{1/2}$ before saturation, using $\Rp = p L < \ell$ denoting a fraction of $L$. This estimate assumes that the backbone filament is absorbed on the cylindrical surface, which is largely true because of a strong bending stiffness due to side loops. 

\subsubsection{In presence of crowders}
The dependence of the number of helical turns on the cell length, however, shows a very different behavior in presence of crowders.  The compression due to crowders maintains the helical morphology even in the longest cells. As the chromosome relaxes the longitudinal stress by opening up in longer cells,  the number of helical turns $n_t$ decreases a little to saturate eventually~(see Fig.\ref{ch_helicity}($c$)\,). The saturation value is determined by the crowder density, lower the density the chromosome opens up more easily with lesser number of turns.

\section{Positioning and Dynamics}
In this section, we discuss a central positioning of the chromosome and the associated sub-diffusive motion. 
\subsubsection{Positioning}
Simulations show a rather precise positioning of the center of mass of the model chromosome to the center of the cell volume, a result that is in good agreement with experiments on live {\em E. coli} cells~\cite{Wu2018}.  In small cells, due to large inter-monomer repulsion the chromosome remains pressed against cell walls. The resultant stress balance causes a central positioning of the chromosome center of mass. As the cell grows, chromosomal compression due to confinement relaxes, however, osmotic pressure due to crowders starts to play an important role. In a live cell, proteins constituting cytosol are produced around the chromosome, via transcription and translation, without any directional bias. 
In our model, as the cell length increased we inserted a spatially symmetric cloud of crowders with a probability proportional to Boltzmann weight around the chromosome, to maintain the crowder density. 
The crowder-mediated stress balance between the model chromosome and cell walls maintains the central location of chromosome in the long cells (e.g., see Fig.\ref{fig:config1}($b$)). Finally, large kinematic barrier in narrow confinement also suppresses fluctuations, leading to an extremely slow dynamics that also contributes towards the observed positioning, to which we return in Sec.\ref{sec:dyn}. 

Note that the chromosome positioning within a given cell length relies on the {\em memory} of central localization in the smaller cell, and the symmetric introduction of crowders around the chromosome. The difference in crowder density between the cell center and cell walls controls the central localization. This kinematic mechanism is maintained by regular insertion of new crowders followed by cell growth, and is distinct from a simple depletion effect in thermal equilibrium. Pure equilibrium depletion, in contrast, would push the polymer against one corner of the cell, opening up larger space for crowders to increase the overall configurational entropy. 

\begin{figure}[t] 
\begin{center}
\includegraphics[width=8.6cm]
{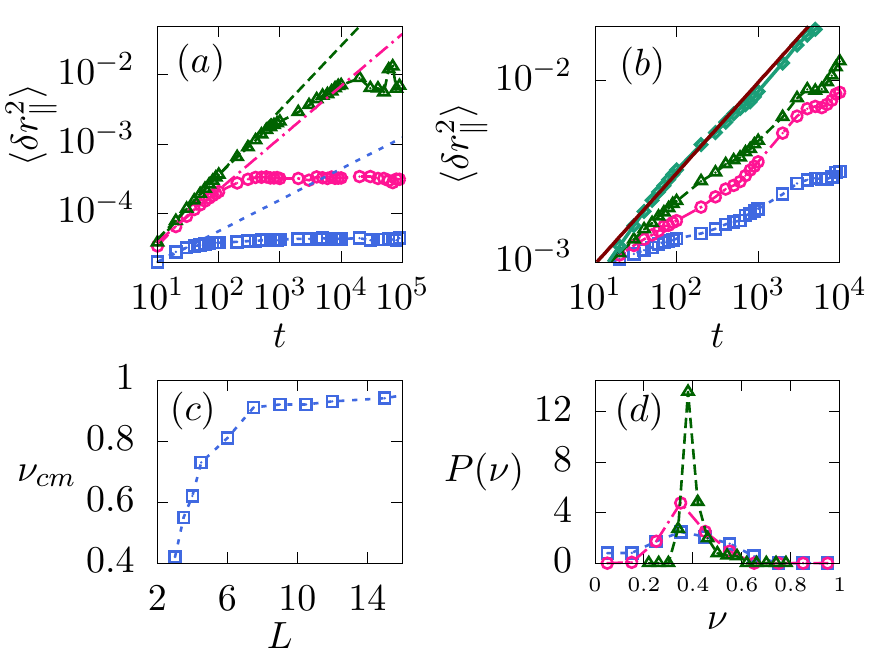}
\caption{(Color online) Mean squared displacement $\la \d r_\parallel^2 \ra$ as a function of time $t$ at $\r D^3=212$ for the center of mass of the chain  and individual monomers show anomalous diffusion $\sim t^\nu$ before confinement controlled saturation. ($a$)\,$\la \d r_\parallel^2 \ra$  at cell lengths $L=3\, (\Box),\, 6\, (\bigcirc),\, 15\, (\bigtriangleup)$. The lines denote power laws $t^{\nu_m}$ with $\nu_m=0.45\, (\Box), 0.75\,(\bigcirc),\, 0.93 (\bigtriangleup)$. ($b$)\,$\la \d r_\parallel^2\ra$ for four different monomers in a cell of length $L=3$. The line denotes power law $t^{0.5}$, characteristic of single file diffusion.  ($c$)\,The exponent corresponding to the center of mass $\nu_{cm}$ grows and saturates with cell length $L$. ($d$)\,Probability distribution $P(\nu)$ of exponent $\nu$ characterizing  MSD of individual loci calculated at cell lengths $L=3\,(\bigcirc),\, 15\,(\bigtriangleup)$ showing broad distributions with peak positions shifting slightly from $\nu = 0.35$($\bigcirc$) to $0.38$ ($\bigtriangleup$). All lengths are expressed in unit of $D$, and time in unit of $\t$.
}
\label{fig:dyn}
\end{center}
\end{figure}

\subsubsection{Dynamics}
\label{sec:dyn}
To analyze the dynamics of the chromosome we study the mean squared displacement (MSD) $\la \d r_\parallel^2 \ra$ of individual loci and the center of mass. The simulation results for these two quantities at $\r D^3=212$ are shown in Fig.s~\ref{fig:dyn}($a$) and ($b$) respectively. As expected, $\la \d r_\parallel^2 \ra$ saturates at long time due to confinement. However, even for longest cells of $L=30\,D$, the amplitude of fluctuations remain small,  $ \la \d r_\parallel^2 \ra^{1/2} < 2D$ (data not shown). This happens even though the chromosomal extension saturates to $\approx 6.5\,D$, in principle, leaving a ``free cytosolic length'' of $23.5\,D$ available for exploration, showing clearly that fluctuations are suppressed by the crowders. 

The short time scale dynamics before saturation can be described by an approximate anomalous diffusion $\la \d r_\parallel^2 \ra \sim t^\nu$ where $\nu$ grows from $\nu \approx 0.45$ at $L=3 D$ to saturate to $\nu \approx 0.93$ by $L=10.5\,D$ (Fig.~\ref{fig:dyn}($c$)). Experiments on non-replicating {\em E. coli} showed similar variation in exponents for center of mass diffusion~\cite{Wu2018}. In absence of crowders, the chromosome center of mass performs simple diffusion before approaching saturation (See Appendix-\ref{app:msd}). The effective anomalous diffusion is thus controlled by the crowders. Fluctuations of individual loci, on the other hand, shows a peaked distribution of $\nu$. The shape of the distribution remains relatively broad in strong confinement~($L=3\,D$), and gets into a narrower and sharply peaked one as the confinement decreases~($L=15\,D$). Fig.~\ref{fig:dyn}($d$) shows the peak position remains largely unaltered, and shifts from $\nu=0.35$ to $0.38$ as cell length increases from $3D$ to $15D$. 
Previous experiments on {\em E. coli} and {\em Caulobacter} showed similar dynamic behavior for chromosomal loci with a distribution of $\nu$ having a maximum near $\nu \approx 0.4$~\cite{Weber2010}. 
Earlier theoretical work described such sub-diffusion in terms of viscoelasticity of medium~\cite{Lampo2015,Weber2010}. 
Our simulations suggest that confinement and osmotic pressure due to depletants are enough to produce such an effective anomalous diffusion.

In a live cell one expects non-equilibrium activity to impact dynamics non-trivially. Although bacteria do not have cytoskeletal motor proteins, treadmilling of cytosolic elements, e.g., TubZ~\cite{Margolin2007}, may provide active energy input. In fact, infrequent super-diffusive bursts of fluctuations in DNA loci have been observed {\em in vivo}~\cite{Javer2014}. Nevertheless, the exponent $\nu$ characterizing predominant fluctuations in live bacteria matches well with our equilibrium simulations, suggesting that such activity can on average be interpreted in terms of an enhanced effective temperature.

\section{Discussion}
Using a simple polymer model, we have shown how cellular confinement and the presence of cytosolic crowders alone can potentially lead to the formation of a compact nucleoid structure of a bacterial chromosome,  which occupies a distinct sub-volume of the cell, without requiring a membrane-bound compartment. We have also shown how the confinement and cytosolic crowders together determine the emergent size, morphology, positioning and dynamics of the chromosome. 

Our results can be translated in terms of real length and time scales by using physiological chromosome length and cytosolic viscosity of specific cell types. For example, interpreting the simulated chain as the 4.6 Mbp long circular DNA of {\em E. coli},
requires the bond length $b \approx 115\,$bp\,=\,39\,nm, the bead size $\s=35.81\,$nm. Thus the $62$ bead side loops turn out to be 7.12 kbp long, which is close to the maximum in loop size distribution presented in Ref.~\cite{Postow2004}. 
The diameter $D=26.67\,\s$ is equivalent to $0.95\,\mu$m, a number close to typical {\em E. coli} cell diameter of $0.8\,\mu$m. We assumed the length to change from $3\,D$ to $30\, D$ that stands for a range of 2.85 to 28.5\,$\mu$m, to examine the impact of longitudinal confinement. 
While the wild type cell length varies between 2 to 4\,$\mu$m, suppression of chromosome replication and cell division by suitable genetic modification and/ or antibiotics allows one to probe such larger length scales to uncover effect of longitudinal confinement on chromosome~\cite{Wu2018}. 
Indeed, our prediction that such experiments on confined growing {\em E. coli} bacteria with fixed diameter, will show a slow increase of chromosomal size saturating near a cell length of $L=15\,\mu$m, agrees well with experiments on live cells~\cite{Wu2018}. The precise value of saturated length does depend on the cytosolic molecular crowding, which we assume to be maintained by ongoing protein production at a rate proportional to the increase in cell length.

The non-additive crowders used in our simulations are a proxy for all molecular species that are big enough to be sterically hindered by the chromosome to contribute to entropic depletion forces. This includes protein molecules with radius of gyration of a few nm, to large protein complexes like ribosomes of size $\sim 20$- $30\,$nm. For DNA, the depletion effect was observed {\em in vitro}  using synthetic crowders of a few nm radius of gyration~\cite{Pelletier2012, Zhang2009}, and recent experiments on live cells showed expulsion of ribosomes outside of {\em E. coli} nucleoid~\cite{Chai2014}.  

The viscosity of eukaryotic cells is known to be $10^2 - 10^3\eta_w$ where $\eta_w=0.001\,$Pa-s, the viscosity of water~\cite{Wirtz2009,Milo2015}. 
In absence of cytoplasmic molecular motors, the dense suspension of bacterial cytosol is expected to have even higher viscosity. 
Using viscosity determined for {\em E. coli} cytosol $\eta=17.5\,$Pa-s~\cite{Kalwarczyk2012}, one gets the unit of time  $\t =1.78\,$s. This leads to an estimate of center of mass diffusivity of chromosome $\approx 2.92 \times 10^{-7}\, \mu {\rm m}^2/$s, in the longest cells of $L=28.5\,\mu$m where the exponent denoting anomalous diffusion $\nu$ approaches unity. Such a small diffusivity contributes to the stable central localization of the chromosome.

The  non-linear growth of chromosome size with cell size is reliably described by an analytical form obtained from mean field theory. We observed a non-monotonic variation of chromosome  extension at small cell lengths, which we have shown to be due to a radial to longitudinal reordering of crowders with respect to the chromosome near the cell poles.  The cylindrical confinement induces a helicoid morphology of the chromosome, which is further stabilized by the osmotic pressure due to molecular crowders. The number of helical turns depends on the crowder density. 

In summary, our predictions for chromosome size and morphology appear to compare well with experiments on rod-shaped bacteria~\cite{Wu2018, Fisher2013, Berlatzky2008}, indicating that we capture a possible physical mechanism behind such robust organization. The dynamics of individual loci and center of mass of the polymer also shows good agreement with that measured for chromosomes in live bacteria\cite{Wu2018, Weber2010}.

We should, however, stress once more that the calculations presented here are equilibrium in nature, with slow dynamics due to strong kinematic barriers arising from confinement and crowders. However, live cells, by definition, are in an out-of-equilibrium state. Thus, the good agreement that we find between our estimate of sub-diffusion exponents with those observed in experiments suggests that such activity in bacteria can possibly be interpreted in terms of an effective equilibrium temperature. 

Finally, our predictions about structure, e.g., chromosome size, shape in terms of number of helical turns, and dynamics can in principle be verified from super-resolution microscopy of live cells, like in Ref.~\cite{Fisher2013}. They may also be applicable to soft matter experiments on synthetic structured polymers in nano-confinement~\cite{Zhang2009}.

\section*{Acknowledgements}
The simulations were performed on SAMKHYA, the high performance computing facility at IOP, Bhubaneswar.
The work of BMM is part of the research programme of the Netherlands Organisation for Scientific Research (NWO). DC acknowledges SERB, India, for financial support through grant number EMR/2016/001454. 
We thank Cees Dekker and Fabai Wu for discussions on related experimental work.

\appendix

\section{Chromosome size in absence of crowders}
\label{no_crowder}

\begin{figure}[t] 
\begin{center}
\includegraphics[width=8.6cm]{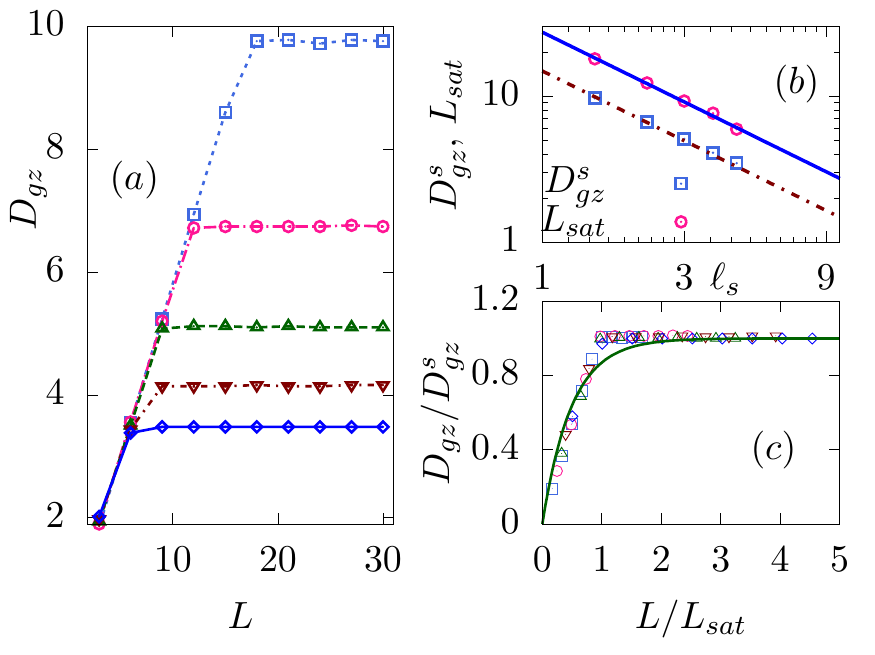}
\caption{(Color online) ($a$)\,Longitudinal size of chromosome $D_{gz}$ increases and saturates 
as a function of cell length $L$.  
All lengths are expressed in terms of cell diameter $D$. 
The symbols $\Box$, $\bigcirc$, $\bigtriangleup$, $\bigtriangledown$ and $\diamond$ show results for chains with side-loop sizes $\ell_s/D=1.5,\, 2.25,\, 3,\, 3.75,\,4.5$, respectively. 
($b$)\,The saturation size $D_{gz}^s$ ($\Box$) and cell lengths $L_{sat}$ ($\bigcirc$) at which saturation occurs decrease with $\ell_s$.  
Both the data sets show scaling form $1/\ell_s$, shown by brown dash-dotted line ($D_{gz}^s$) and blue solid line ($L_{sat}$).
($c$)\,Data collapse of chromosome size versus cell length. An exponential saturation, $1 - \exp(-x/x_s)$ is shown by the line, where $x=L/L_{sat}$ and $x_s = 0.49$.}
\label{scaling}
\end{center}
\end{figure}
In this section we discuss the dependence of chromosome size and shape on the degree of confinement, in absence of crowders.
Apart from end-to-end extension, chromosome size may be determined in terms of its radius of gyration. The cylindrical confinement breaks the isotropy of the system. The chromosome size along the length of the cell 
$D_{gz} = 2 \times R_{gz}$, where $R_{gz}$ denotes the component of the radius of gyration tensor along the $z$- axis, the long axis of the cylinder. Like $\Rp$,
$D_{gz}$ increases linearly with the cell length to finally saturate~(Fig.~\ref{scaling}($a$)). The saturation length $D^s_{gz}$ and the cell-length at which the saturation occurs $L_{sat}$, depends on the length of side-loops $\ell_s$~(see Fig.~\ref{scaling}($a$)~), incorporated via the effective Gaussian core interaction between loops as discussed in the main text.  
 
Both $D^s_{gz}$ and $L_{sat}$ show approximate $1/\ell_s$ scaling (Fig.~\ref{scaling}($b$)), and their behavior can be understood using the following argument. Given that the chain length $\ell=(n_s +1) n_m \s$ is constant, the backbone length $n_m \s \sim \ell/ n_s$. The radius of gyration of side-loops follows the Flory scaling $n_s^{3/5} \s$. Repulsion between side-loops renders an effective bending rigidity to the backbone. This gives a length scale over which the polymer bending remains suppressed. Interpreting this as a persistence length $\l \sim n_s^{3/5} \s$ of the backbone, we can estimate the effective size of the polymer. The configuration of such a semiflexible chain strongly confined in narrow cylinder may be viewed as consisting of {\em rays} of length $\l$ getting reflected between two walls separated by a distance $D$~\cite{Odijk1983}. Between two reflections, this covers a longitudinal distance $\sqrt{\l^2 - D^2}$. One gets $n_m \s/\l$ number of such reflections. Thus the total length of such a chain in an infinitely long cylinder would be
\bea
D^s_{gz} &=& {\sqrt{\l^2 - D^2}}\, \f{n_m \s}{\l} \nn\\
&=& \sqrt{1-\le( \f{D}{n_s^{3/5} \s}\ri)^2}\, \f{\ell}{n_s} \sim \f{1}{\ell_s},
\eea   
the length to which the polymer size saturates (Fig.\ref{scaling}($b$)).

In Fig.\ref{scaling}($c$) we show the data collapse obtained by plotting $D_{gz}/D^s_{gz}(\ell_s)$ against $L/L_{sat}$. 
This suggests a scaling  function
\bea
D_{gz} = D^s_{gz} f(L/L_{sat})
\approx \f{\ell}{n_s} f \left( \f{n_s L}{\s }\right),
\eea
shown in the figure.

\section{Constant number of crowders}
\label{fixed_Nd}

\begin{figure}[t] 
\begin{center}
\includegraphics[width=8cm]{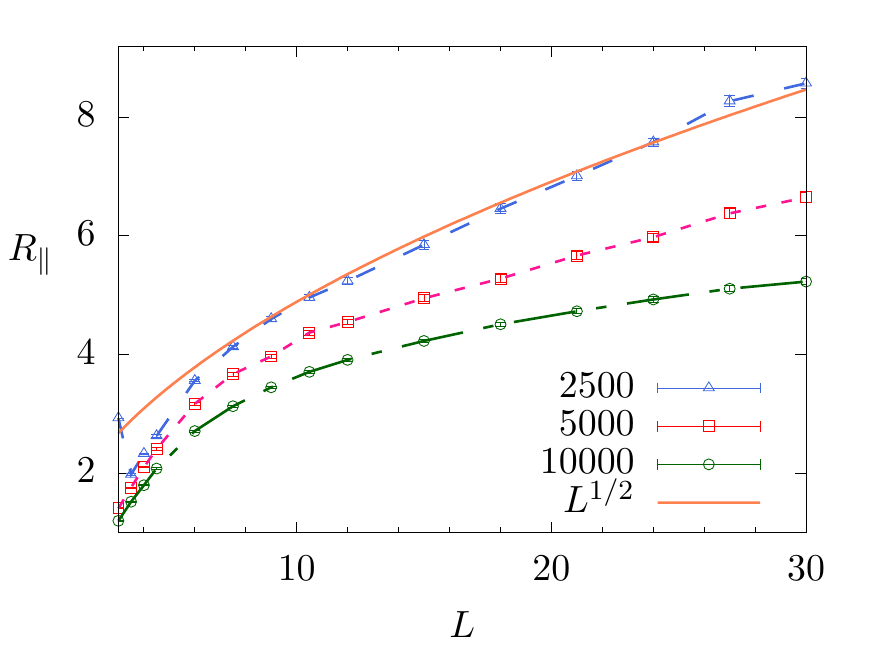}
\caption{(Color online) Longitudinal size of the chromosome at constant number of depletants $N_d=2500,\, 5000,\, 10000$. Lengths are expressed in units of $D$. Before saturation, $\Rp \sim L^{1/2}$ as shown by the red solid line.}
\label{fig:const_nd}
\end{center}
\end{figure}
\begin{figure}[t]
\begin{center}
\includegraphics[width=8.6cm]
{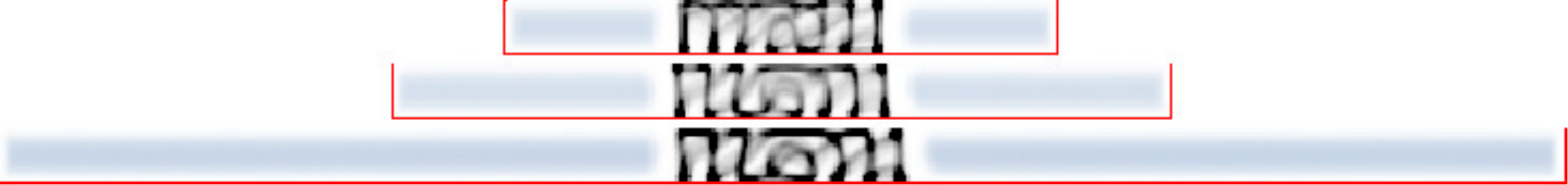}
\caption{(Color online) Chromosome structure with increasing cell length, at a fixed crowder density $\r D^3=1060$. 
The plots from top to bottom depict local density of chromosome and crowders for increasing cell lengths
$L/D=10.5,\, 15,\, 30$.  
}
\label{fig:configs_ro1060}
\end{center}
\end{figure}
\begin{figure}[h] 
\begin{center}
\includegraphics[width=8cm]{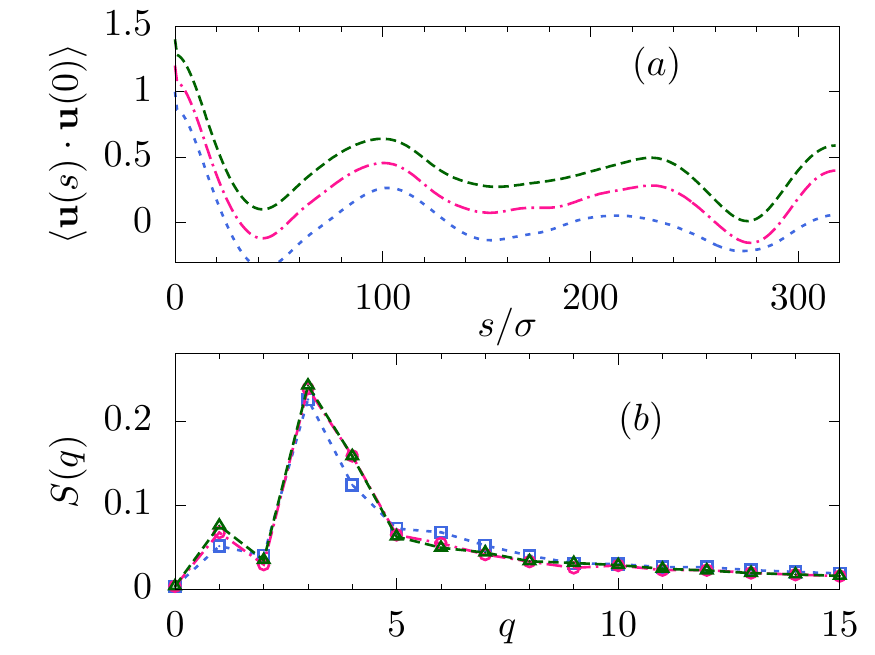}
\caption{(Color online) ($a$) Local tangent  correlation function $\la {\bf u}(s)\cdot {\bf u}(0)\ra$ along the chromosome backbone from $s=0$ to $\ell_m/2$ at crowder density $\r D^3=1060$. The plots are  for cell lengths $L/D=10.5,\, 15,\, 30$ denoted by blue dashed line, magenta dash- dotted line, green big-dashed line, respectively. Each graph corresponding to a larger $L$-value is shifted upwards 
by an additive factor $0.2$ for better visibility. ($b$)~Corresponding structure factors at $L/D=10.5\,(\Box),\, 15\,(\bigcirc),\, 30\, (\bigtriangleup)$ capture the periodicity in correlation, with the  largest peak position at $q = 3$ denoting the number of turns $n_t=6$ in all the three cases shown here. }
\label{fig:tangent3}
\end{center}
\end{figure}
\begin{figure}[t] 
\begin{center}
\includegraphics[width=8cm]{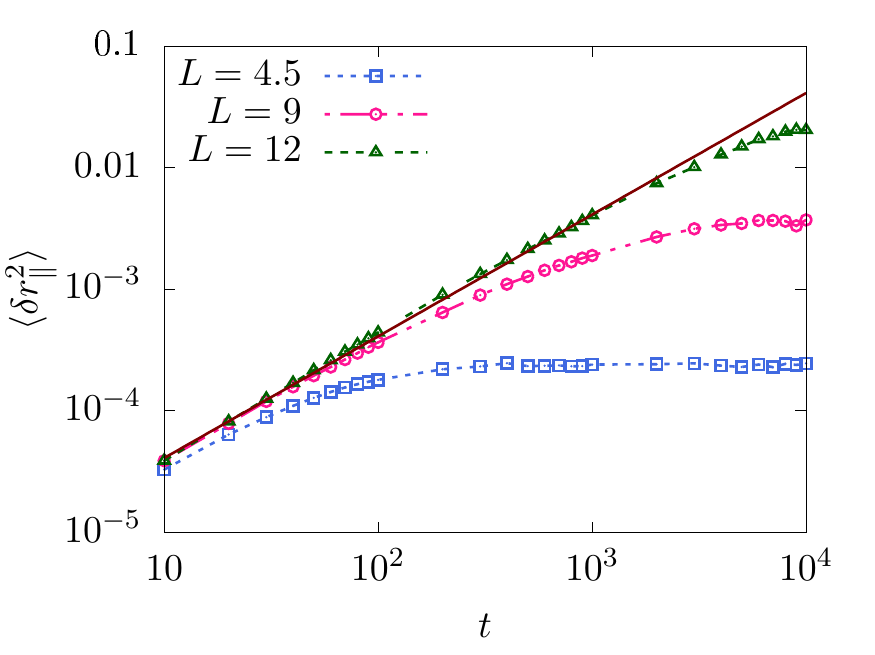}
\caption{(Color online) Longitudinal mean squared displacement $\la \d r_\parallel^2 \ra / D^2$ of center of mass as a function of time $t/\t$  at cell lengths $L/D=4.5,\,7,\, 12$. The straight line is a plot of  $\la \d r_\parallel^2 \ra = 2 {\cal D} t$ with 
${\cal D}=1.7 \times 10^{-6}\, D^2/\t$.
}
\label{diff}
\end{center}
\end{figure}

In wild type cells, the cytosolic density remains largely unchanged to maintain metablic activity as the cell grows. However, if a cells behavior is controlled by mutation or otherwise to stop replication and cell division~\cite{Wu2018}, in long enough cells, presumably, a single chromosomal copy may not be able to maintain unaltered cytosolic density. 
In such cases, the cell may get into a regime of constant number of crowders, leading to a reduction in density with cell growth. Here, we show how chromosome size would change if the number of cytosolic crowders remains constant in a growing cell.

Fig.~\ref{fig:const_nd} shows the variation of chromosome extension with cell length, for fixed number of crowders $N_d$. With increasing $L$ the crowder density drops, finally to vanish in the limit $L \to \infty$. In large $L$ limit, thus,  the chromosome size saturates to complete expansion $\Rp = (b/a)^{1/3} \sim N D^{-2/3}$ consistent with confining diameter $D$. As predicted by Eq.~\ref{eq:rp}, it  increases as $\Rp \sim L^{1/2}$ before that limit is reached (see discussion in Sec.\ref{mft}).

\section{Chromosome shaped by crowders}
\label{app:morpho}

Here we show dependence of chromosomal morphology on cell length, for a high density of crowders $\r D^3=1060$. As we discussed in the main text, the cytosol, modeled as non-additive crowders, phase segregate towards cell walls and generate a osmotic pressure stabilizing the helical morphology of chromosomes inside cylindrical confinement, even for the longest cells (Fig.\ref{fig:configs_ro1060}). This is in stark contrast to chromosome in absence of crowders where in longest cells helical turns do open up.

The chromosomal shape is quantified in terms of the local tangent- tangent  correlation function $\la {\bf u}(s) \cdot {\bf u}(0)  \ra$, and the corresponding structure factor. We show here results for these two quantities at $\r D^3= 1060$.  Fig.~\ref{fig:tangent3}($a$) shows clearly how the increasing number of turns with cell length is captured by increasing oscillations in the correlation function. The corresponding peak of structure factor at $q=q_p$ in Fig.~\ref{fig:tangent3}($b$) gives the number of helical turns $n_t=2 q_p$.

\section{Dynamics in absence of crowders}
\label{app:msd}
The dynamics of chromosome in absence of crowders  shows a simple diffusive scaling, before mean squared displacement saturates. This behavior is shown in Fig.\ref{diff}. 
Longer cells show diffusive scaling $\la \d r_\parallel^2 \ra \sim t$ over longer time scales. 



\begin{thebibliography}{10}

\bibitem{Wang2013}
X. Wang, P.~M. Llopis, and D.~Z. Rudner, Nature Reviews Genetics {\bf 14},  191
   (2013).

\bibitem{Fisher2013}
J.~K. Fisher, A. Bourniquel, G. Witz, B. Weiner, M. Prentiss, and N. Kleckner,
  Cell {\bf 153},  882  (2013).

\bibitem{Worcel1972}
A. Worcel and E. Burgi, J. Mol. Biol. 71: {\bf 71},  127  (1972).

\bibitem{Holmes2000}
V. Holmes and N. Cozzarelli, Proc. Natl. Acad. Sci. {\bf 97},  1322  (2000).

\bibitem{Zimmerman2006}
S.~B. Zimmerman, Journal of Structural Biology {\bf 156},  255   (2006).

\bibitem{Brocken2018}
D.~J. Brocken, M. Tark-Dame, and R.~T. Dame, Curr. Opin. Syst. Biol. {\bf 8},
  137  (2018).

\bibitem{Goloborodko2016}
A. Goloborodko, M.~V. Imakaev, J.~F. Marko, and L. Mirny, Elife {\bf 5},  1
  (2016).

\bibitem{Song2015}
D. Song and J.~J. Loparo, Trends Genet. {\bf 31},  164  (2015).

\bibitem{Delius1974}
H. Delius and A. Worcel, Cold Spring Harb. Symp. Quant. Biol. {\bf 38},  53
  (1974).

\bibitem{Kavenoff1976}
R. Kavenoff and B. Bowen, Chromosoma {\bf 59},  89  (1976).

\bibitem{Trun1998}
N. Trun and J. Marko, ASM News {\bf 64},  276  (1998).

\bibitem{Postow2004}
L. Postow, C.~D. Hardy, J. Arsuaga, and N.~R. Cozzarelli, Genes Dev. {\bf 18},
  1766  (2004).

\bibitem{Le2013}
T.~B.~K. Le, M.~V. Imakaev, L.~A. Mirny, and M.~T. Laub, Science (80-. ). {\bf
  342},  731  (2013).

\bibitem{Marbouty2014}
M. Marbouty, A. Cournac, J.~F. Flot, H. Marie-Nelly, J. Mozziconacci, and R.
  Koszul, Elife {\bf 3},  e03318  (2014).

\bibitem{Berlatzky2008}
I.~A. Berlatzky, A. Rouvinski, and S. Ben-Yehuda, Proc. Natl. Acad. Sci. U. S.
  A. {\bf 105},  14136  (2008).

\bibitem{Weber2010}
S.~C. Weber, A.~J. Spakowitz, and J.~A. Theriot, Phys. Rev. Lett. {\bf 104},
  27  (2010).

\bibitem{Kuwada2013}
N.~J. Kuwada, K.~C. Cheveralls, B. Traxler, and P.~A. Wiggins, Nucleic Acids
  Res. {\bf 41},  7370  (2013).

\bibitem{Javer2014}
A. Javer, N.~J. Kuwada, Z. Long, V.~G. Benza, K.~D. Dorfman, P.~a. Wiggins, P.
  Cicuta, and M.~C. Lagomarsino, Nat. Commun. {\bf 5},  3854  (2014).

\bibitem{Lampo2015}
T.~J. Lampo, N.~J. Kuwada, P.~A. Wiggins, and A.~J. Spakowitz, Biophys. J. {\bf
  108},  146  (2015).

\bibitem{Parry2014}
B. Parry, I. Surovtsev, M. Cabeen, C. O'Hern, E. Dufresne, and C.
  Jacobs-Wagner, Cell {\bf 156},  183  (2014).

\bibitem{Reiss2011}
P. Reiss, M. Fritsche, and D.~W. Heermann, Phys. Rev. E {\bf 84},  051910
  (2011).

\bibitem{Chaudhuri2012}
D. Chaudhuri and B.~M. Mulder, Physical Review Letters {\bf 108},  268305
  (2012).

\bibitem{Chaudhuri2018}
D. Chaudhuri and B.~M. Mulder,  in {\em Bacterial Chromatin}, edited by R.~T.
  Dame (Springer Protocols, Humana Press, New York, 2018), Chap.~Molecular
  Dynamics Simulation of a Feather-Boa Model of a Bacterial Chromosome, p.\
  403.

\bibitem{Odijk1998}
T. Odijk, Biophys. Chem. {\bf 73},  23  (1998).

\bibitem{Woldringh1995}
C. Woldringh, FEMS Microbiol. Lett. {\bf 131},  235  (1995).

\bibitem{Mondal2011}
J. Mondal, B.~P. Bratton, Y. Li, A. Yethiraj, and J.~C. Weisshaar, Biophys. J.
  {\bf 100},  2605  (2011).

\bibitem{Chai2014}
Q. Chai, B. Singh, K. Peisker, N. Metzendorf, X. Ge, S. Dasgupta, and S.
  Sanyal, J. Biol. Chem. {\bf 289},  11342  (2014).

\bibitem{Nevo-Dinur2011}
K. Nevo-Dinur, A. Nussbaum-Shochat, S. Ben-Yehuda, and O. Amster-Choder,
  Science (80-. ). {\bf 331},  1081  (2011).

\bibitem{Wu2018}
F. Wu, P. Swain, L. Kuijpers, X. Zheng, K. Felter, M. Guurink, D. Chaudhuri, B.
  Mulder, and C. Dekker, bioRxiv  https://doi.org/10.1101/348052  (2018).

\bibitem{Kremer1990}
K. Kremer and G.~S. Grest, J. Phys. Condens. Matter {\bf 2},  SA295  (1990).

\bibitem{Weeks1971}
J.~D. Weeks, J. Chem. Phys. {\bf 54},  5237  (1971).

\bibitem{Bolhuis2001}
P.~G. Bolhuis, A.~A. Louis, J.~P. Hansen, E.~J. Meijer, and I. Introduction,
  Chemical Physics {\bf 114},  4296  (2001).

\bibitem{Narros2013}
A. Narros, A.~J. Moreno, and C.~N. Likos, Macromolecules {\bf 46},  9437
  (2013).

\bibitem{Dijkstra1998}
M. Dijkstra, R. van Roij, and R. Evans, Phys. Rev. Lett. {\bf 81},  2268
  (1998).

\bibitem{Kuhlman2012}
T.~E. Kuhlman and E.~C. Cox, Mol. Syst. Biol. {\bf 8},  1  (2012).

\bibitem{Binder1997}
K. Binder, Reports Prog. Phys. {\bf 60},  487  (1997).

\bibitem{Widom1963}
B. Widom, J. Chem. Phys. {\bf 39},  2808  (1963).

\bibitem{Cloizeaux1975}
J. des Cloizeaux, Journal de Physique {\bf 36},  281  (1975).

\bibitem{Daoud1975}
M. Daoud, J.~P. Cotton, B. Farnoux, G. Jannink, G. Sarma, H. Benoit, C.
  Duplessix, C. Picot, and P.~G. de~Gennes, Macromolecules {\bf 8},  804
  (1975).

\bibitem{Grosberg1982}
A.~Y. Grosberg, P.~G. Khalatur, and A.~R. Khokhlov, Die Makromolekulare Chemie,
  Rapid Communications {\bf 3},  709  (1982).

\bibitem{Gennes1979}
P.-G. de~Gennes, {\em Scaling concepts in polymer physics} (Cornell University
  Press, Ithaca, 1979).

\bibitem{Jun2006}
S. Jun and B.~M. Mulder, Proc. Natl. Acad. Sci. U. S. A. {\bf 103},  12388
  (2006).

\bibitem{Limbach2006}
H.-J. Limbach, A. Arnold, B.~A. Mann, and C. Holm, Comput. Phys. Commun. {\bf
  174},  704  (2006).

\bibitem{Margolin2007}
W. Margolin, Curr Biol. {\bf 17},  R633  (2007).

\bibitem{Pelletier2012}
J. Pelletier, K. Halvorsen, B.-Y. Ha, R. Paparcone, S.~J. Sandler, C.~L.
  Woldringh, W.~P. Wong, and S. Jun, Proc. Natl. Acad. Sci. {\bf 109},  E2649
  (2012).

\bibitem{Zhang2009}
C. Zhang, P.~G. Shao, J.~A. van Kan, and J.~R.~C. van~der Maarel, Proc. Natl.
  Acad. Sci. {\bf 106},  16651  (2009).

\bibitem{Wirtz2009}
D. Wirtz, Annu. Rev. Biophys. {\bf 38},  301  (2009).

\bibitem{Milo2015}
R. Milo and R. Phillips, {\em Cell biology by the numbers} (Garland Science,
  New York, 2015).

\bibitem{Kalwarczyk2012}
T. Kalwarczyk, M. Tabaka, and R. Holyst, Bioinformatics {\bf 28},  2971
  (2012).

\bibitem{Odijk1983}
T. Odijk, Macromolecules {\bf 16},  1340  (1983).

\end{thebibliography}
\end{document}